# Semicylindrical microresonator: excitation, modal structure, and Q factor

## H. HAROYAN, H. PARSAMYAN, T. YEZEKYAN, KH. NERKARARYAN*

*Department of Radiophysics, Yerevan State University, A. Manoogian 1, Yerevan 0025, Armenia*
*Corresponding author: knerkar@ysu.am*

The semicylindrical microresonator with relatively simple excitation with plane wave is studied. The resonator is formed on the base of the dielectric / metal / dielectric structure, where the wave energy penetrates into resonator through a thin metal layer and stored in a semicylindrical dielectric with high permittivity. The proposed microresonator combines features of Fabry-Perot and Whispering gallery mode resonators. Dependence of radiation losses on the radius and materials are estimated by theoretical analysis, while excitation by a plane wave is shown via numerical analysis. The quality *Q*-factor of the resonator can achieve up to $10^4$, at radius of a semicylinder of several microns.

http://dx.doi.org/10.1364/AO.99.099999

## 1. INTRODUCTION

Efficient and low-loss optical coupling to high quality (*Q*) factor Whispering Gallery Mode (WGM) microresonators [1] is important for a wide range of applications include frequency [2,3] and soliton mode-locked microcombs [4–8], bio and nano-particle sensors [9–11], cavity optomechanical oscillators [12], Raman lasers [13], and quantum optical devices [14,15]. Dielectric resonators on various platforms (e.g. silica, silicon nitride, lithium niobate) have been monolithically integrated with on-chip waveguides, achieving loaded *Q*-factor as high as $10^8$ [16–22]. Usually, to achieve phase-matched and mode-matched evanescent wave efficient coupling, it is necessary to use a host material of resonator with relatively low refractive index compared to those of standard waveguide coupling materials. In addition, fiber tapers as unclad waveguides are quite brittle and applicable only for resonators, refractive index of which is close to the refractive index of fiber. At the same time waveguide coupling requires tens or hundreds micrometer diameters of resonators (or coupling region) to achieve effective phase-matching. Thus, excitation of high refractive index WGM resonators with diameter of few micrometers is quite challenging [22–24]. Meanwhile, for Biosensing purposes, binding of single virions is observed from discrete changes in the resonance frequency of a WGM excited in a microcavity. It is shown, that the magnitude of the discrete wavelength-shifted signal can be sufficiently enhanced by reducing the microsphere size [9,25]. On the other hand, the effective control of the light wave as a rule is realized using materials with a large refractive index.

Hence, it is interesting to consider microresonators with high refractive index of host material and dimensions close to the exciting wavelength.

In this paper a simple structure of microresonator with easy coupling method with incident plane wave is proposed. The resonator is based on dielectric / metal / dielectric structure, where the wave energy is stored in the semicylindrical dielectric with a high refractive index (see Fig.1). The exciting plane wave normally incidents on a dielectric medium ($\varepsilon_d$) with plane-parallel boundaries and penetrates into the semicylinder ($\varepsilon_s$) through a thin metal layer ($\varepsilon_m$).

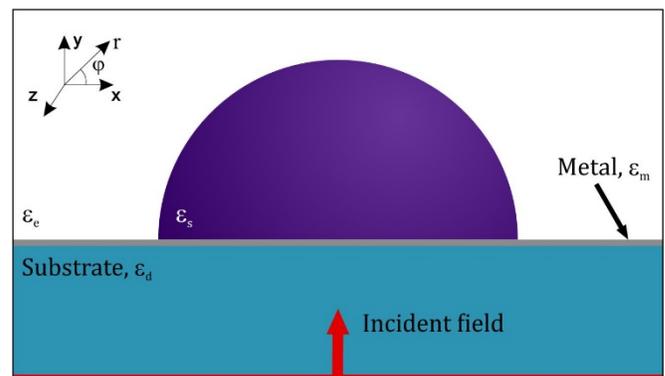

**Fig. 1.** The cross section of the structure of semicylindrical microresonator. $\varepsilon_d = 2.25$, $\varepsilon_e = 1$ are dielectric permittivities of the substrate (SiO$_2$ Silica) and the surrounding medium (Air), respectively. As a metal and semicylinder medium Ag and GaAs were used correspondingly.

Proposed resonator combines properties of Fabry-Perot resonator, where the input of wave energy is carried through mirrors, and cylindrical resonator, where Whispering-gallery ($WG_{m,\ell}$) modes with azimuthal $m$ and radial $\ell$ mode numbers are formed.

Here, the possibility of simple input and output of radiation is combined with the possibility of using the unique properties of an evanescent wave on cylindrical surface of a dielectric. Although values of the *Q*-factor of the proposed resonator are less than that of planar microdisk and microring resonators, this is quite sufficient for a wide range of investigations.

The investigation is divided into two parts focused on two aspects of functioning of the microcavity. At first, an analytical determination of dependence of the radiated field from curved cylindrical surface of WGM resonator on the radius and the ratio of the dielectric permittivities of the semicylinder and surrounding is realized. This defines conditions under which the radiation part of the *Q*-factor is negligible. Secondly, numerical study is performed to clarify the possibility of the excitation of semicylindrical microresonator directly by a plane wave which determines optimal values of the *Q*-factor and the role of Joule losses in the metallic layer.

## 2. THEORY

Let's start with a theoretical study of WGM formation in the proposed structure, to identify (determine) the range of values of the radius of a half-cylinder, where the radiation losses of the microcavity can be neglected.

Assume, $\varepsilon_d$, $\varepsilon_m$, $\varepsilon_s$ and $\varepsilon_e$ are the permittivity's of the substrate, the metal layer, the semicylinder and the surrounding medium respectively, $R$ is the radius of the semicylinder, and $h$ is the thickness of the metal layer (see Fig. 1). We use the cylindrical coordinate system ($r$, $\varphi$, $z$), where $z$ is directed along the axis of the semicylinder. Let's consider the case when the electromagnetic wave, on the side of the dielectric substrate, incidences normally on to the structure (Fig. 1), with the electric field polarized along the axis of the semicylinder ($z$ axis).

The WGMs formed in the semicylindrical resonator are also solutions of the Helmholtz equation in a cylindrical geometry with additional boundary condition: on the metal surface $E_z$ component of electric field is equal to zero. The Helmholtz equation written in cylindrical coordinates for the axial field of a TM WGM is:

$$\left( \frac{\partial^2}{\partial r^2} + \frac{1}{r}\frac{\partial}{\partial r} + \frac{1}{r^2}\frac{\partial^2}{\partial \varphi^2} + k_{s,e}^2 \right) E_z(r,\varphi,t) = 0, \quad (1)$$

where $k_{s,e} = \sqrt{\varepsilon_{s,e}} \frac{\omega}{c} = \frac{2\pi}{\lambda_{s,e}}$

Here $\omega$ is the angular frequency, $\lambda_{s,e} = \lambda_0 / \sqrt{\varepsilon_{s,e}}$ are wavelengths in the semicylinder and surrounding medium, respectively ($\lambda_0$ is vacuum wavelength). This equation can be simplified via the method of separation of variables by which it is split into two equations for radial and azimuthal components. These modes consist of azimuthally propagating fields guided by total internal reflection at the dielectric interface and optical interference which prevents the field from penetrating inward beyond a fixed radius.

The appropriate solutions for the radial field dependence both interior ($r \leq R$) and exterior ($r > R$) to the semicylinder are [26]:

$$E_z(r,\varphi,t) = A J_m(k_s r) \cdot \sin(m\varphi) \cdot \exp(i\omega t) \quad (2)$$

at $r \leq R$, $0 < \varphi < \pi$,

$$E_z(r,\varphi,t) = B H_m^{(1)}(k_e r) \cdot \sin(m\varphi) \cdot \exp(i\omega t) \quad (3)$$

at $r > R$, $0 < \varphi < \pi$.

Here $J_m(k_s r)$ and $H_m^{(1)}(k_e r)$ are the Bessel and Hankel functions respectively, $m$ is the azimuthal number ($m$=1, 2, ...), A and B are unknown constants.

Here we assume that on the surface of the metal layer the tangential component of the electric field is equal to zero. As mentioned before, the aim of this section is to theoretically evaluate radiation losses from curved cylindrical surface of the resonator; hence for this purpose aforementioned assumption is quite acceptable. However, hereinafter within numerical analysis to determine conditions for a microcavity excitation by a plane wave, this assumption certainly will not be used.

From the continuity of the tangential components of the field it can be obtained [26,27]:

$$\sqrt{\varepsilon_s} \frac{J'_m(k_s R)}{J_m(k_s R)} = \sqrt{\varepsilon_e} \frac{\left[ H_m^{(1)}(k_e R) \right]'}{H_m^{(1)}(k_e R)} \quad (4)$$

$$A J_m(k_s R) = B H_m^{(1)}(k_e R) \quad (5)$$

Family of WGMs naturally come up from the solution of Eq. (4), which are indexed by two mode indices $m$ (azimuthal mode number), $\ell$ (the radial mode number) and characterized by resonant frequencies $\omega_{m,\ell}$ [28]. Here, as well as in the Eqs. (2) and (3) only explicit dependence on $m$ is present, although the $\ell$ is defined by the order ($m$), argument of the Bessel function and radius of the semicylinder. For a fixed wavelength (of an existing mode) and radius, $\ell$ indicates number of extremums of Bessel function of order $m$ till the boundary. It is worth to mention, that from stated boundary conditions (tangential component of the electric field is equal to zero on the metal surface) directly follows that azimuthal mode number is odd (otherwise, surface currents will appear). Note, that for a given $m$, the numbering $\ell$ is performed with increasing resonant frequency. On the other hand, the Eq. (5) gives relation between the inner and outer fields of resonator. In further analysis we are focusing on the WGM of $\ell$ =1 due to lowest mode volume, which, in fact, is much more valuable from practical point of view. Particularly, in several experiments with virus-sized polystyrene nanoparticles it comes out, that there is a mechanism for increasing signal by limiting modal volume [25].

The first maximum of $J_m(k_s r)$ function, when $\ell$ =1 ($WG_{m,1}$) and $m \gg 1$ is located at $k_s r \approx m$. Hence, in the discussed case $\varepsilon_s \gg \varepsilon_e$ when $k_e R \ll m$ for estimations we can use following approximations [29]:

$$H_m^{(1)}(k_e R) \approx -i \frac{(m-1)!}{\pi} \left( \frac{2}{k_e R} \right)^m \quad (6)$$

$$J_m(m) \approx \frac{2^{1/3}}{3^{2/3} \Gamma(2/3)} \frac{1}{m^{1/3}} \quad (7)$$

$$m! \approx m^m \exp(-m) \sqrt{2\pi m} \quad (8)$$

Finally, the relation between constants of inner and outer fields has following form:

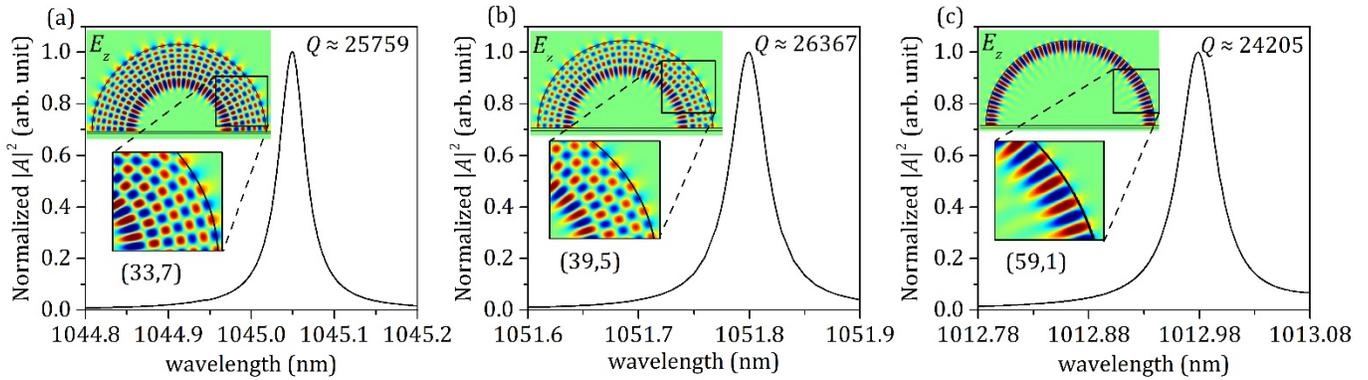

**Fig. 2.** Distributions of the electric filed $E_z$ component amplitudes in particular modes, resonance curves, and $Q$ factors: (a) $m=33, \ell=7$ ($WG_{33,7}$), $\lambda_0 = 1045.05$ nm (b) $m=39, \ell=5$ ($WG_{39,5}$), $\lambda_0 = 1051.8$ nm and (c) $m=59, \ell=1$ ($WG_{59,1}$), $\lambda_0 = 1012.98$ nm. The model parameters are: $\varepsilon_d = 2.25$, $\varepsilon_e = 1$, $R = 3$ μm, $h = 80$ nm. As a metal and material of semicylinder silver (Ag) and GaAs were used correspondingly.

$$B \approx iA \sqrt{\frac{\pi m^{1/3}}{2^{1/3}}} \frac{1}{3^{2/3}\Gamma(2/3)} \left(\frac{e}{2}\sqrt{\frac{\varepsilon_e}{\varepsilon_s}}\right)^m \quad (9)$$

where $m \approx 2\pi R/\lambda_s$.

Note that dependence of the $|B|/|A|$ on the semiclyinder radius $R$ (which is derived from Eq. 9 and the relation between $m$ and $R$) is exponentially decaying function. Since the radiative part of the $Q$-factor of the resonator is proportional to the ratio $|A|^2/|B|^2$, therefore from Eq. (9) it is possible to determine conditions when radiation from the curved boundary is negligible (which is possibly realized), and the $Q$-factor of resonator is mainly determined by the radiation from metal layer and Joule's losses.

## 3. NUMERICAL ANALYSIS AND DISCUSSION

Now, let us consider the problem of the excitation of WGM in a semicylindrical microresonator by a plane wave. Numerical analysis based on Finite Element Method is carried out for the following values of the parameters: $\varepsilon_d = 2.25$, $\varepsilon_e = 1$, $R = 3$ μm, $h = 80$ nm, the values of $\varepsilon_m$ (as a metal we use silver) and $\varepsilon_s$ (as a semiconductor we use GaAs) have been chosen according to references [30] and [31] respectively.

The characteristic images for the electric field distributions ($E_z$ component) of modes with a relatively high $Q$ are presented in Fig. 2. The $Q$-factor of the resonator was determined by equation $Q \approx \lambda_p / \Delta\lambda$ where $\lambda_p$ and $\Delta\lambda$ are the peak wavelength and the full width of the half maximum respectively. $\Delta\lambda$ was determined from the curve of dependence of the square of the amplitude of the field on the wavelength. In the investigated case like in the case of traditional WGM resonators the $Q$-factor of resonator is proportional to the radius, the bigger radius the higher $Q$-factor. It should be noted that the calculations carried out according to Eq. (2) and the results obtained by numerical calculations perfectly correspond to each other.

To optimize metal thickness, it is interesting to investigate the dependencies of electric field inside the resonator and $Q$-factor on metal thickness $h$. To carry out the dependence of the square of electric field amplitude ($|A|^2$) on the metal thickness, we fixed the resonant wavelength at 1012.98 nm and changed the metal thickness in range of 50 nm to 120 nm. After that we derived the maximum value of the $|A|^2$ within the semicylinder. Secondly, to derive the dependence of the $Q$-factor of $WG_{59,1}$ on the metal thickness, we changed the metal thickness and determined the $Q$-factor by the same way as for Fig. 2. Numerical calculations show that $Q$-factor increases by increasing the thickness of the metal layer and saturates approximately started from $h = 80$ nm (see Fig. 3). Meanwhile, the square of the amplitude of the z-component of the electric field (or, in essence, the wave energy stored in the microcavity) reaches a maximum at $h = 65$ nm, after which it decreases monotonically. Such a seeming discrepancy can be explained only by a rather rapid decrease in the total (radiation and Joule) losses of the microcavity.

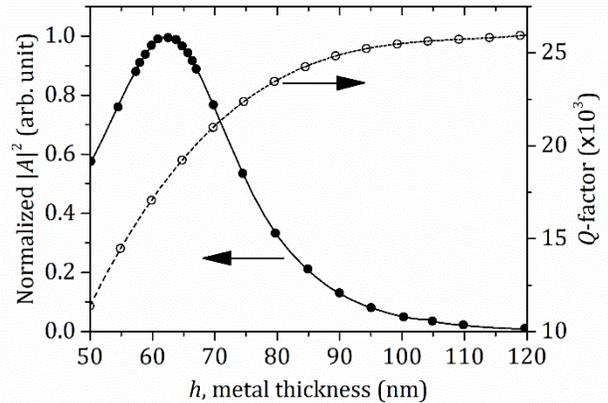

**Fig. 3.** Dependences of the normalized square of electric field amplitude $|A|^2$ (left axis) and $Q$-factor (right axis) on the thickness $h$ of metal layer for $WG_{59,1}$ mode.

The point is that by increasing the thickness of the metal layer, the value of electric component of the standing wave on the surface of the metal layer promptly tends to zero.

In this context it is important to indicate the dielectric constants of Ag ($\varepsilon_m$) and GaAs ($\varepsilon_s$) at resonant wavelengths: $\varepsilon_m = -52.068 + 0.57727i$ and $\varepsilon_s = 12.2444$ at $\lambda_0 = 1012.98$ nm ($WG_{59,1}$), $\varepsilon_m = -55.7 + 0.59707i$ and $\varepsilon_s = 12.132$ at

$\lambda_0 = 1045.05$ nm (WG$_{33,7}$), $\varepsilon_m = -56.481 + 0.60124i$ and $\varepsilon_s = 12.1097$ at $\lambda_0 = 1051.8$ nm (WG$_{39,5}$).

Thus, in order to maximize the wave fields in the microcavity it is necessary to choose a certain thickness of the metal layer, while for maximizing $Q$-factor another value should be chosen. Hence an optimal thickness of metal layer must be picked to satisfy the demands of relatively high values of wave field amplitude in microcavity and at the same time preserving reasonable values of $Q$-factor. Note, that $h$=80 nm ensures both high values of $Q$-factor and relatively high values of the field inside the resonator. This creates favorable conditions for the registration of the process and has an important role in terms of the vastness of the applications.

For parameters used in the numerical analysis $|B|/|A|$ derived from Eq. 9 is approximately $10^{-8}$, hence the radiative losses from the curved boundary are negligible. Relatively high value of $Q$-factor is ensuring extreme sensitivity of the wave energy stored in the resonator from the number of parameters. In particular, a noticeable change of the wave energy stored in the resonator with a change in the refractive index $n_s = \sqrt{\varepsilon_s}$ of the semicylinder by $\Delta n_s \approx 10^{-4}$ opens up wide possibilities for light control by using various nonlinear and electro-optical effects (see Fig. 4, bottom axis). It is noteworthy that the wave energy is weakly dependent on the refractive index of the surrounding medium.

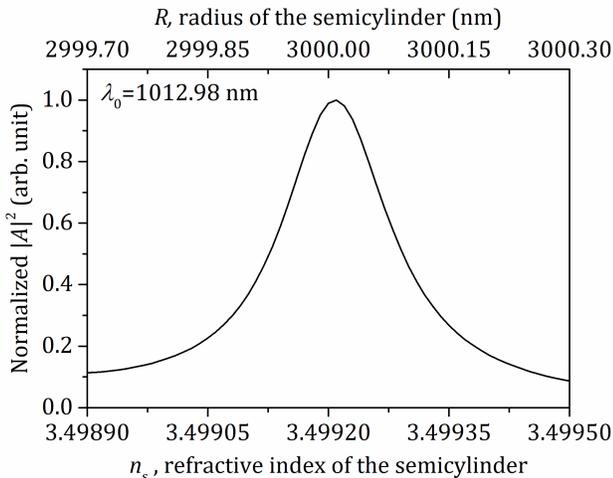

**Fig. 4.** Dependence of the square of electric field amplitude $|A|^2$ on the refractive index of medium filled in half-cylinder $n_s$ (bottom axis) and on the radius of half-cylinder $R$ (top axis) for WG$_{59,1}$ resonant mode. The metal thickness $h = 80$ nm, for surrounding medium $\varepsilon_e = 1$. Note that top and bottom axes are independent.

Another interesting feature is that the energy stored in resonator significantly depends on semicylinder radius (see Fig. 4, top axis).

From numerical calculations it follows that in the resonant regime stored energy significantly changes by varying the radius of semicylinder, even when changing by decimal part of nanometer. This circumstance gives base to conclude that the investigated system is able to register the presence of a molecular thin layer absorbed by the semicylindrical resonator.

## 4. CONCLUSION

Thus, based on the dielectric / metal / insulator structure, an optical microresonator can be designed where the wave energy is stored in a semicylindrical dielectric with a relatively high permittivity. Here, the exciting wave falls from the side of another dielectric and penetrates into the microcavity through a thin metal layer, thereby creating favorable conditions for an easily coupling. The proposed microresonator combines the properties of Fabry-Perot resonator, where the input of the wave energy is carried out through mirror (metal layer), and the cylindrical resonator, where whispering-gallery modes are formed.

Dependence of radiation losses on the radius and materials are estimated by theoretical analysis, while excitation by a plane wave is shown via numerical analysis.

Calculations show that the $Q$-factor can achieve up to $10^4$, which is competitive with another whispering-gallery mode resonators with sizes of several micrometers. The high values of $Q$ ensure the extreme sensitivity of the wave energy stored in the resonator from the number of parameters. In particular, a noticeable change of the wave energy stored in the resonator by the alteration of the refractive index of the semicylinder by $\Delta n_s \approx 10^{-4}$ opens up wide possibilities for light control using various nonlinear and electrooptical effects. The energy stored in resonator significantly depends on the semicylinder radius, which makes it possible to use as a sensor for detection of the absorbed molecular thin layers. The simple way to experimentally study the process of accumulating wave energy is to measure the output power from the cylindrical surface by evanescent coupling. In the case of the formation of a resonance mode, the output power should drastically increase. Another way is the measurement of the reflected field disruption from metal layer.

**Funding.** State Committee of Science of the Republic of Armenia (15T-1C146), MES RA-BMBF STC Foundation (GE-027).